\documentclass[]{aa}
\usepackage{graphicx}
\usepackage{txfonts}
\usepackage{natbib}

\bibpunct{(}{)}{;}{a}{}{,}

\newcommand\bb[1] {   \mbox{\boldmath{$#1$}}  }
\newcommand\del{\bb{\nabla}}
\newcommand\bcdot{\bb{\cdot}}
\newcommand\btimes{\bb{\times}}

\begin{document}

\title{MHD simulations of the
  magnetorotational instability  in a shearing
  box with zero net flux. II. The effect of transport coefficients}
\author{ S\'ebastien Fromang \inst{1}, John Papaloizou \inst{1}
  , Geoffroy Lesur \inst{2} and Tobias Heinemann \inst{1}}

\offprints{S.Fromang}

\institute{Department of Applied Mathematics
and Theoretical Physics, University of Cambridge, Centre for
Mathematical Sciences, Wilberforce Road, Cambridge, CB3 0WA, UK \and
Laboratoire d'astrophysique, UJF CNRS, BP 53, 38041
Grenoble Cedex 9 \\ \email{S.Fromang@damtp.cam.ac.uk}}

\date{Accepted; Received; in original form;}

\label{firstpage}

\abstract
{}
{We study the influence of the choice of
  transport coefficients (viscosity and resistivity) on MHD
  turbulence driven by the magnetorotational instability (MRI) in
  accretion disks.} 
{We follow the methodology described in paper I: we adopt 
  an unstratified shearing box model and focus on the case where the
  net vertical magnetic flux
  threading the box vanishes. For the most part we use the 
    operator split code
  ZEUS, including explicit transport coefficients in the
  calculations. However, we also compare our results
  with those obtained using other  algorithms (NIRVANA, the PENCIL code and
  a spectral code) to demonstrate both the convergence of our results and their
  independence of the numerical scheme.}
{We find that small scale dissipation affects the saturated state of MHD
  turbulence. In agreement with  recent similar numerical simulations
  done in the presence of a net vertical magnetic
  flux, we find that turbulent activity (measured by the rate of
  angular momentum transport) is an increasing function of the
  magnetic Prandtl number $Pm$ for all values of the Reynolds number
  $Re$ that we investigated. We also found that turbulence disappears
  when the Prandtl
  number falls below a critical value $Pm_{c}$ that is apparently a
  decreasing function of $Re$. For the limited region of parameter
  space that can be probed with current computational resources, we always
  obtained $Pm_{c}>1$.}
{We conclude that the magnitudes of the transport coefficients are
  important in determining the properties of MHD turbulence in
  numerical simulations in the shearing box with zero net flux,
  at least for Reynolds numbers and magnetic
  Prandtl numbers that are such that transport is not dominated by
  numerical effects and thus can be probed using current computational
  resources.}

\authorrunning{S.Fromang et al.}
\titlerunning{MHD turbulence in accretion disks. II. Transport coefficients}
\maketitle

\section{Introduction} 
\label{intro}

To date, the magnetorotational instability (MRI) is believed to be the
most likely cause of anomalous angular momentum transport in accretion
disks \citep{balbus&hawley98}. The results of numerous numerical
simulations carried out over the last 15 years indicate that the MRI
results in MHD turbulence that transports  angular momentum outwards
with rates that seem, depending on the net magnetic flux present, to
be compatible to order of magnitude with those estimated  from the
observations. Most of these studies were local MHD numerical
simulations of a shearing box performed using finite difference
methods. Unfortunately because of the limited computational resources
available during the 90's, these early calculations were restricted to
low and moderate resolutions, having at most $64$ grid cells per disk
scale height \citep{hawleyetal95,hawleyetal96,flemingetal00,sanoetal04}.

In paper I \citep{Fropap07}, we used one of these  numerical codes,
ZEUS \citep{hawley&stone95}, to study the convergence of the results
as the grid resolution
is increased. We explored resolutions ranging  from $64$ to $256$ grid
cells per scale height and  concentrated on the special case for which
the net vertical flux threading the box vanishes. We found that the
angular momentum transport, measured using the standard $\alpha$
parameter,  decreases linearly with the grid spacing as the resolution
increases. In the best resolved simulations, we obtained
$\alpha=10^{-3}$, which amounts to a decrease of about one order of
magnitude when compared to the earlier estimates derived from the
first simulations of this problem \citep{hawleyetal95}. This situation
comes about because significant flow energy  and dissipation occurring
at the grid scale  affects the numerical results, at least for the
resolutions that can currently be achieved.

Since the diffusive transport 
and  dissipation that plays an important role in determining the 
outcome is
purely numerical in ZEUS, we concluded that explicit physical 
dissipation, both viscous and resistive, should be properly included
 in the simulations for them to show numerical convergence and thus have physical
significance. The purpose of this paper is to investigate the
effect of specified transport coefficients
  on the saturated state of MRI driven MHD turbulence.

We note that this issue has recently been considered by
\citet{lesur&longaretti07} for cases for which a nonzero net vertical 
flux threads the disk. They found that both the viscosity $\nu$ and
the resistivity $\eta$ affect the amount of angular momentum
transported by MHD turbulence, in such a way that $\alpha$ increases
with the magnetic Prandtl number $Pm=\nu/\eta$. 
One of the aims of
this paper is to
investigate whether such a relation exists in the absence of
net flux and to quantify the precise rate of angular momentum
transport for that case as a function of the transport coefficients.

The plan of the paper is as follows. In section~\ref{run_setup}, we
describe our numerical setup and the models we
simulated. Section~\ref{fiducial_sec} focuses on the properties of one
of these models. Specifically, we demonstrate  that
for the simulations performed with ZEUS, numerical dissipation does not
strongly affect the results when explicit transport coefficients of sufficient
magnitude are included.
This is done by using  Fourier analysis as
introduced in paper I, and by comparing the simulation  results
obtained using a  variety of  numerical schemes. Having validated our
approach, we turn in
section~\ref{parameter_sec} to a more systematic exploration of the
parameter space. We recover the correlation between $\alpha$ and $Pm$
mentioned above and surprisingly identify a regime of parameters in
which MHD turbulence decays. Finally, we discuss our results, their
limits and their astrophysical implications in section~\ref{conclusion}.

\section{Model properties}
\label{run_setup}

As in paper I, we work in the framework of the shearing box
 model \citep{goldreich&lyndenbell65}. The standard ideal MHD
 equations are modified to account for small scale dissipation
 resulting from finite and constant viscosity $\nu$ and resistivity
 $\eta$. For clarity they are recalled below. 
\begin{eqnarray}
\frac{\partial \rho}{\partial t} + \del \bcdot (\rho \bb{v})  &=&  0 \,\label{contg} , \\
\rho \frac{\partial \bb{v}}{\partial t} + \rho ( \bb{v} \bcdot \del )
\bb{v}   + 2 \rho
\bb{\Omega} \times \bb{v}&=&\frac{ (\del \btimes \bb{B})
\btimes \bb{B}}{4\pi} -\del P +\nabla \bcdot \bb{T} , \\
\frac{\partial \bb{B}}{\partial t}  &=&  \del \btimes ( \bb{v} \btimes
\bb{B}  - \eta \del \btimes \bb{B} ) \, \label{induct} .
\label{shearing_sheet_eq}
\end{eqnarray}
As in paper I, $\Omega$ is the Keplerian
angular velocity at the centre of the box, $\rho$ is the gas
density, $\bb{v}$ is the velocity, $\bb{B}$ is the magnetic field and
$P$ is the pressure.  This is related to the gas density and
the constant speed of sound $c_0$ through the isothermal equation of
state $P=\rho c_0^2.$
   We adopt a Cartesian coordinate system $(x,y,z)\equiv (x_1,x_2,x_3)$ with associated unit vectors
  $(\bb{i},\bb{j},\bb{k}).$ 

 Viscosity enters the equations   through
the viscous stress tensor $\bb{T}$ whose components are defined
following \citet{landau&lifchitz59}
\begin{equation}
T_{ik}=\rho \nu \left( \frac{\partial v_i}{\partial x_k} +
\frac{\partial v_k}{\partial x_i} -\frac{2}{3} \delta_{ik} \del \bcdot \bb{v}
\right) \, .
\end{equation}
As in paper I, the simulations are performed within a box  with dimensions
$(L_x,L_y,L_z)=(H,\pi H,H)$, where $H=c_0/\Omega$ is the disk scale
height. Throughout this paper, we used the resolutions
$(N_x,N_y,N_z)=(128,200,128)$ or $(256,400,256)$ depending on the
magnitude of the dissipation coefficients. All the simulations are
initialised with the same  density and  magnetic field as in paper
I. The initial velocity is taken to be given by the 
Keplerian shear with the addition of a random perturbation of small amplitude to
 each component. As in paper I, we measure time in units of
the orbital period $2\pi/\Omega$. 

\begin{table*}[t]
\begin{center}\begin{tabular}{@{}cccccccc}\hline\hline
Model & Resolution & Reynolds number & Pm & $\alpha_{Rey}$ & $\alpha_{Max}$ & $\alpha$ & Turbulence? \\
\hline\hline
128Re800Pm4 & $(128,200,128)$ & 800 & 4 & - & - & - & No \\
128Re800Pm8 & $(128,200,128)$ & 800 & 8 & $5.8 \times 10^{-3}$ & $2.6 \times 10^{-2}$ & $3.1 \times 10^{-2}$ & Yes \\
128Re800Prm16 & $(128,200,128)$ & 800 & 16 & $8.1 \times 10^{-3}$ & $3.6 \times 10^{-2}$ & $4.4 \times 10^{-2}$ & Yes \\
\hline
128Re1600Pm2 & $(128,200,128)$ & 1600 & 2 & - & - & - & No \\
128Re1600Pm4 & $(128,200,128)$ & 1600 & 4 & $1.6 \times 10^{-3}$ & $8.1 \times 10^{-3}$ & $9.7 \times 10^{-3}$ & Yes \\
128Re1600Pm8 & $(128,200,128)$ & 1600 & 8 & $3.3 \times 10^{-3}$ & $1.6 \times 10^{-2}$ & $1.9 \times 10^{-2}$ & Yes \\
\hline
128Re3125Pm1 & $(128,200,128)$ & 3125 & 1 & - & - & - & No \\
128Re3125Pm2 & $(128,200,128)$ & 3125 & 2 & - & - & - & No \\
128Re3125Pm4 & $(128,200,128)$ & 3125 & 4 & $1.6 \times 10^{-3}$ & $7.4 \times 10^{-3}$ & $9.1 \times 10^{-3}$ & Yes \\
\hline
128Re6250Pm1 & $(128,200,128)$ & 6250 & 1 & - & - & - & No \\
128Re6250Pm2 & $(128,200,128)$ & 6250 & 2 & - & - & - & No \\
\hline
128Re12500Pm1 & $(128,200,128)$ & 12500 & 1 & - & - & - & No \\
256Re12500Pm2 & $(256,400,256)$ & 12500 & 2 & $3.2 \times 10^{-4}$ &
$1.3 \times 10^{-3}$ & $1.6 \times 10^{-3}$ & Yes \\
\hline
256Re25000Pm1 & $(256,400,256)$ & 25000 & 1 & - & - & - & No \\
\hline\hline
\end{tabular}
\caption{Properties of the simulations presented in this paper and performed
  with the finite difference code ZEUS. The first column gives the
  label of the model while the resolution $(N_x,N_y,N_z)$ is specified in
  the second column. The third column gives the Reynolds number
  associated with the viscosity~,~$(c_0 H)/\nu),$ and the fourth column gives the
  magnetic Prandtl number $Pm$. The following three columns quantify the
  amount of turbulent activity, when nonzero, through the values of
  $\alpha_{Rey}$, $\alpha_{Max}$ and $\alpha$. Finally, the last
  column describes the outcome, turbulent or not, of each model.}
\label{modes_properties_dissip}
\end{center}
\end{table*}

The parameters for the  simulations we performed with ZEUS are  given in
Table~\ref{modes_properties_dissip}. In section~\ref{code_compar_sec},
we explore the sensitivity of some of these results to the numerical
algorithm by performing a few additional runs using other codes. The
details of these simulations as well as the properties of those codes
are described there. In
Table~\ref{modes_properties_dissip}, we give the label of each model
in the first column and  the resolution $(N_x,N_y,N_z)$ in the
second column. As mentioned above, the models include explicit
viscosity and resistivity,  the values of these  can  be  found
 from the Reynolds number $Re$ (given in the third column) and the magnetic
Prandtl number $Pm$ (given in the fourth column). The former is defined by 
\begin{equation}
Re=\frac{c_0 H}{\nu} \,
\end{equation}
while the latter, already mentioned in the introduction, quantifies the
relative importance of ohmic  and viscous diffusion:
\begin{equation}
Pm=\frac{\nu}{\eta}=\frac{Re_M}{Re} \,
\end{equation}
where   the magnetic Reynolds number $Re_M=c_0H/\eta.$

The  aim of this paper is to study the properties of MHD turbulence in
each of these models. Table~\ref{modes_properties_dissip} summarise
the results by giving, for each of them, the rate of angular momentum transport
that MHD turbulence, when present, generates. This is done through
specifying the Reynolds stress parameter, $\alpha_{Rey},$  the
Maxwell stress parameter $\alpha_{Max}$ and the total stress parameter 
 $\alpha$ in
columns five, six and seven  respectively.
 These represent actual stresses normalised by the initial  thermal
pressure.   They are  defined precisely  in paper I  through Eq.~(6), (7) and
(8). As will be emphasised below, some models fail
to show sustained turbulence. This is why the last column of
Table~\ref{modes_properties_dissip} describes the nature, turbulent or
not, of the flow for each  model.  For cases   in which MHD
turbulence is found to decay, no stress parameters are given.

\section{Re=3125, Pm=4}
\label{fiducial_sec}

\subsection{Simulation set up}
\label{re3125pm4}

In this section, we concentrate on the particular model 
128Re3125Pm4 for which 
$Re=3125$ and $Pm=4$ in order to describe  common features
characterising  simulations of this type.
 Model 128Re3125Pm4 was computed for a duration of 440 orbits
 using ZEUS with a
resolution $(N_x,N_y,N_z)=(128,200,128)$. When using this code,
it is important to be sure that dissipation arising
from the specified diffusion coefficients $\nu$ and $\eta$ dominates 
that due to numerical effects. This is verified in section~\ref{fourier_ana} using
the Fourier analysis  described  in paper I. We also compare the
results from model 128Re3125Pm4 to results obtained from 
similar models computed using other
numerical methods (see section~\ref{code_compar_sec}). This  shows
that the results obtained in the present paper are independent of the
numerical scheme used. But before considering these points in detail, we
first describe the overall properties of model 128Re3125Pm4.

\subsubsection{Simulation properties}

\begin{figure}
\begin{center}
\includegraphics[scale=0.5]{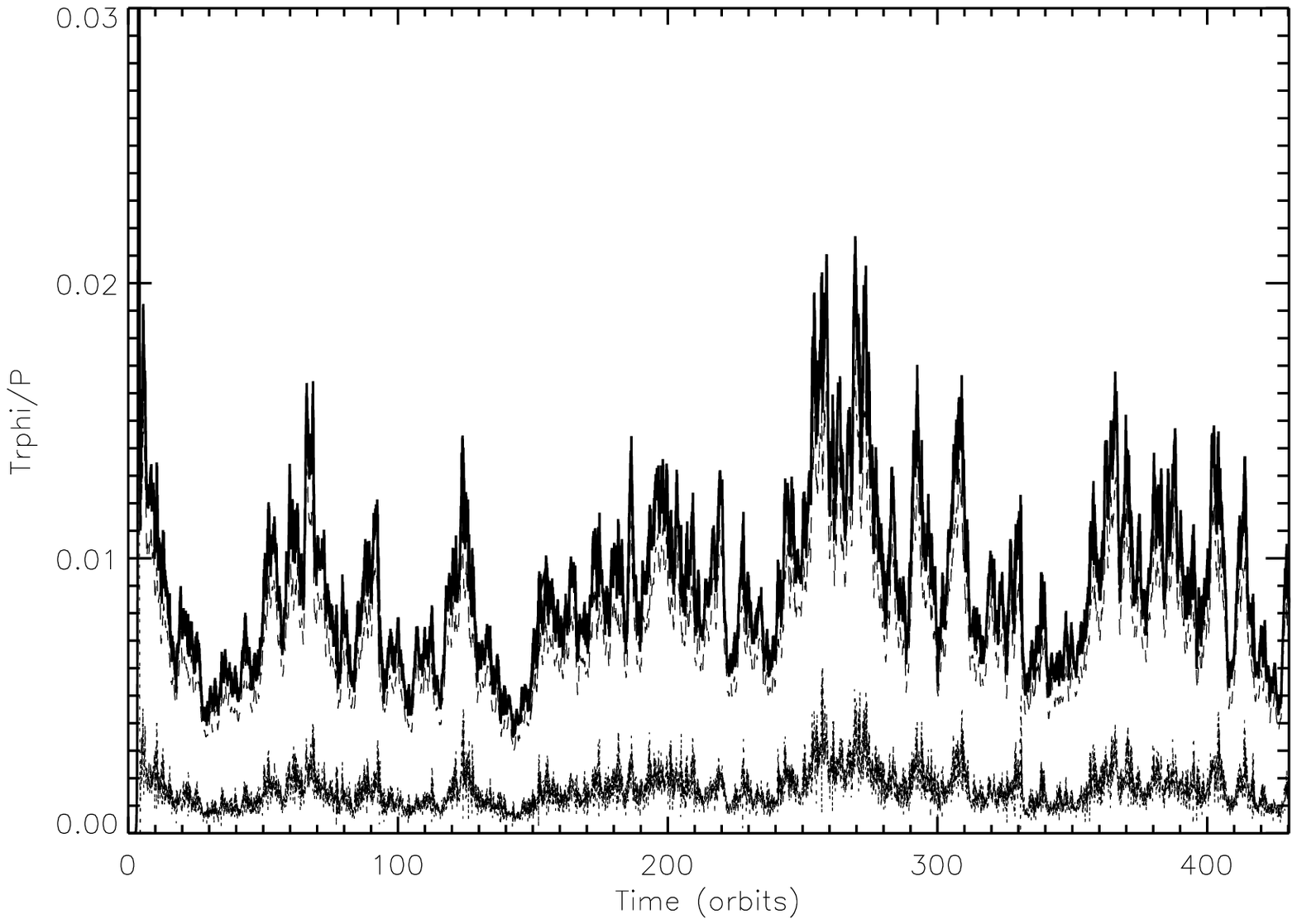}
\caption{Time history of $\alpha_{Re}$ ({\it dotted line}),
  $\alpha_{Max}$ ({\it dashed line}) and $\alpha$ ({\it solid line})
  for model 128Re3125Pr4. From the rate of angular momentum transport
  averaged between $t=40$ and the end of the simulation we obtain
  $\alpha=9.1 \times 10^{-3}$.}
\label{fiducial_transport}
\end{center}
\end{figure}

\begin{figure*}
\begin{center}
\includegraphics[scale=0.4]{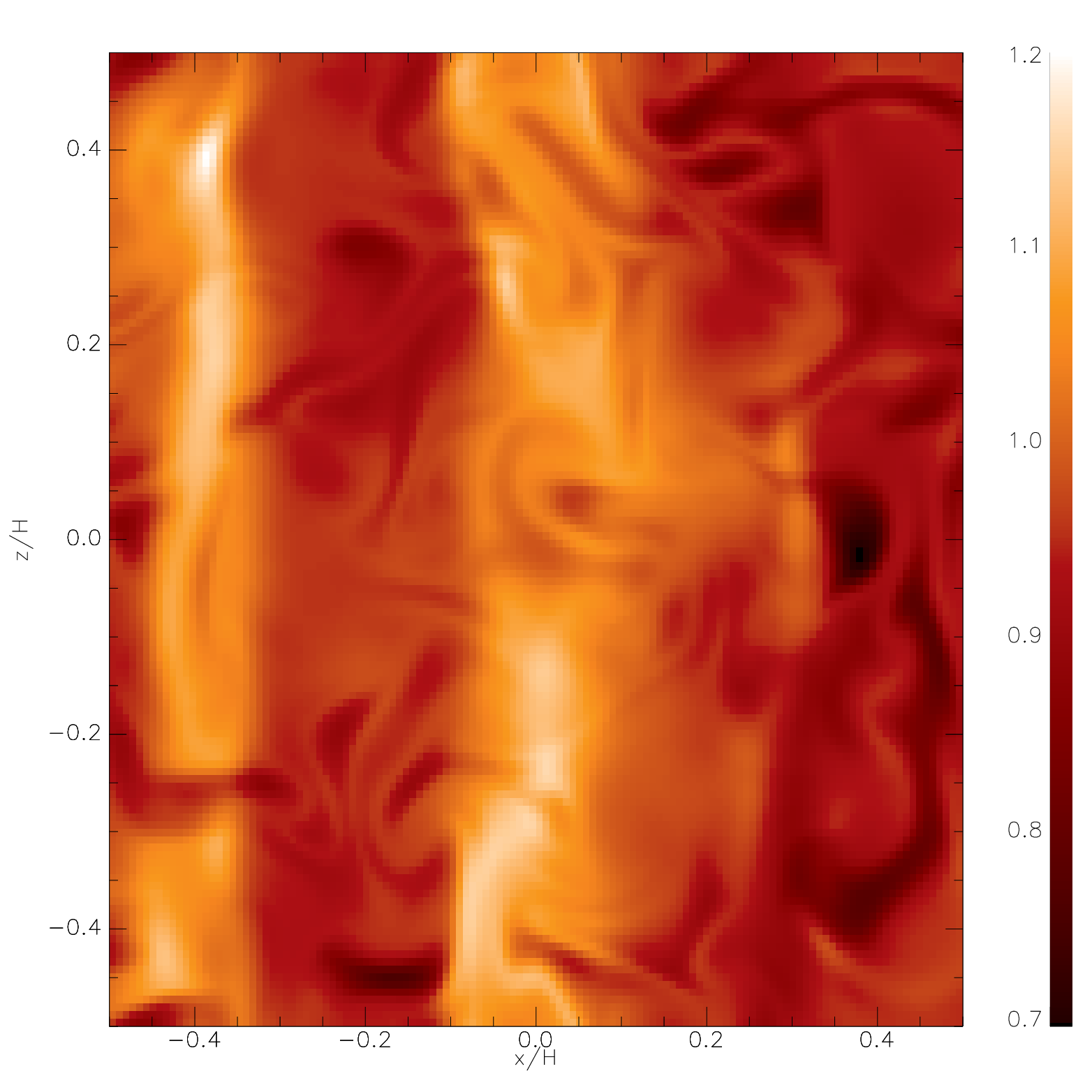}
\includegraphics[scale=0.4]{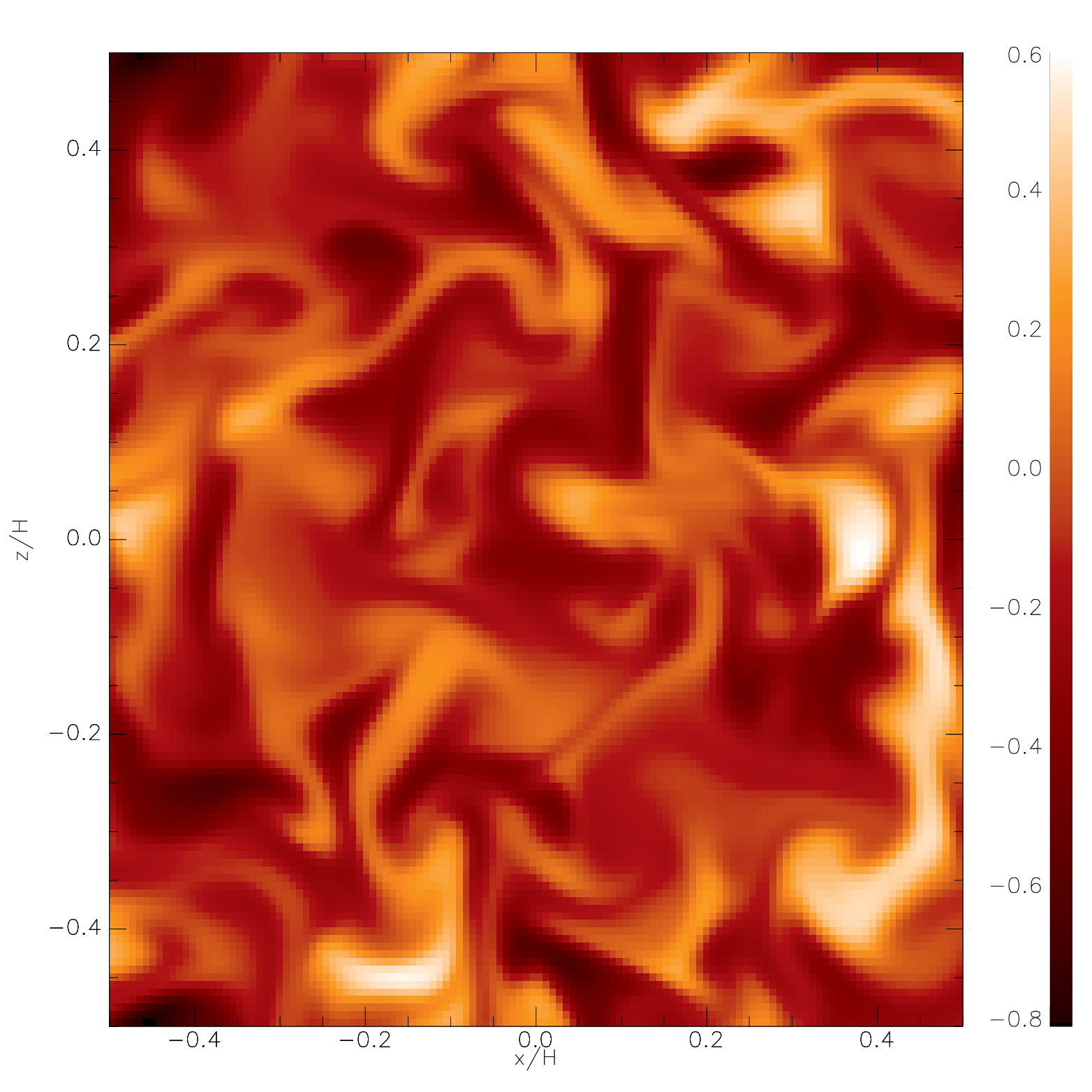}
\caption{Snapshot of the density ({\it left panel}) and of the y
  component of the magnetic field ({\it right panel}) in the $(x,z)$
  plane for model 128Re3125Pm4. In  the former case, the local density is
  normalised by the mean density in the box. In the latter case, the
  magnetic field is normalised by the square root of the mean thermal
  pressure in the box.}
\label{fiducial_snapshots}
\end{center}
\end{figure*}

Despite effects arising from the specification
of nonzero dissipation coefficients, the overall
evolution of model 128Re3125Pm4 is qualitatively very similar to that
found in numerical simulations of MHD turbulence in the shearing
box since the early $1990$'s. 
The MRI destabilises the flow, which
causes the Maxwell and Reynolds
stresses to grow exponentially during the first few orbits.
 When the disturbance reaches a large enough
amplitude, the instability enters the non linear regime, the flow breaks down into MHD
turbulence and eventually settles into a quasi steady state during
which angular momentum is transported outwards.
 The rate of angular
momentum transport is illustrated in figure~\ref{fiducial_transport}
which shows the time history of $\alpha_{Rey}$ ({\it dotted line}),
$\alpha_{Max}$ ({\it dashed line}) and $\alpha$ ({\it solid line}).
 Averaging $\alpha$
in time between $t=40$ and the end of the simulation, we find
$\alpha=9.1 \times 10^{-3}$. As with simulations done without
including explicit dissipation coefficients, the transport is
dominated by the Maxwell stress, which is found to be roughly $5$
times larger than the Reynolds stress. As in paper I, we checked
  that the shearing box boundary conditions did not introduce any
  spurious net flux in the y and z directions. To do so, we measured
the maximum value reached by the mean magnetic field components in the box and
translated their resulting strengths into effective $\beta$. This gave
$\beta_y=5.0 \times 10^{6}$ and
$\beta_z=5.6 \times 10^{6}$ respectively for the azimuthal and
vertical mean components. These values are comparable to the
same quantities obtained in paper I in the absence of dissipation
coefficients and correspond to field strength far too small to play a
role in the model evolution.

In order to illustrate the properties of the flow in this model, in figure~\ref{fiducial_snapshots} we show two
snapshots which represent the density ({\it
  left panel}) and azimuthal component of the magnetic field, $B_y,$ ({\it
  right panel}) in the $(x,z)$ plane at time $t=115.$
   As usual with 
such simulations (whether or not explicit dissipation
coefficients are included), density waves develop and propagate radially in the
disk. They are superposed on smaller scale fluctuations correlated
with the  magnetic field
fluctuations seen on the right side of
Fig.~\ref{fiducial_snapshots}.
 Note, however, the larger characteristic scale
of these 
fluctuations compared to those found in a simulation having the same resolution
 and
no explicit dissipation coefficients (a typical snapshot from such a
simulation is available in the middle panel of Fig.4 in paper I).
 This
is because resistivity and viscosity sets a typical dissipation
length scale that is
larger than a grid cell. This is a first indication that explicit
dissipation  dominates over numerical dissipation in this
simulation. A more quantitative demonstration can be obtained from
 the Fourier analysis approach of paper I. 
 This is considered in the next section.

\subsubsection{Fourier analysis}
\label{fourier_ana}

\begin{figure}
\begin{center}
\includegraphics[scale=0.4]{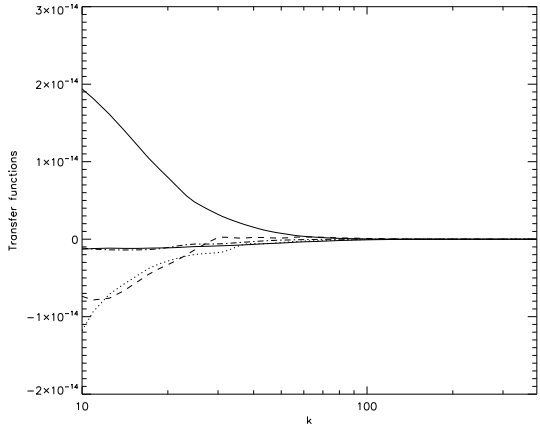}
\caption{Variation of  $S$ ({\it upper solid line}), $T_{bb}$
  ({\it dashed line}), $T_{divv}$ ({\it dotted line}), $T_{bv}$ ({\it
  dotted--dashed line}) and $D_{phys}$ ({\it lower solid line}) as functions of the
   modulus of the wavenumber $k$ for model 128Re3125Pm4. Only $S$ is
always  positive. As explained in the text, this accounts for the rate of 
  toroidal magnetic field generation by the mean shear flow acting
  on the poloidal field.}
\label{transfer_tot_Re3125_Pr4}
\end{center}
\end{figure}

\begin{figure}
\begin{center}
\includegraphics[scale=0.4]{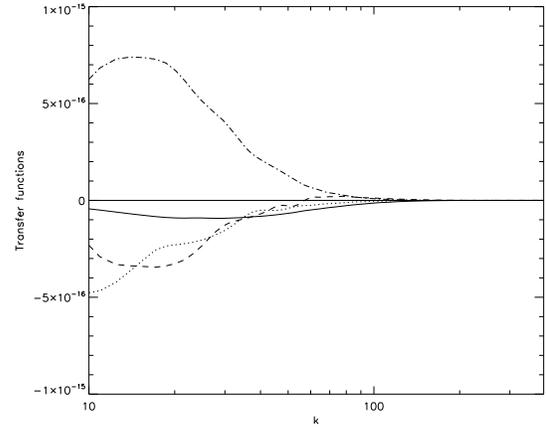}
\caption{As in  figure~\ref{transfer_tot_Re3125_Pr4} but for the
  analogue  of Eq.~(\ref{balance_eq}) derived from the poloidal
  components of the induction equation. This shows that poloidal
  magnetic field energy is created at all scales through field line
  stretching due to the turbulent velocity fluctuations.}
\label{transfer_polo_Re3125_Pr4}
\end{center}
\end{figure}

\begin{figure}
\begin{center}
\includegraphics[scale=0.4]{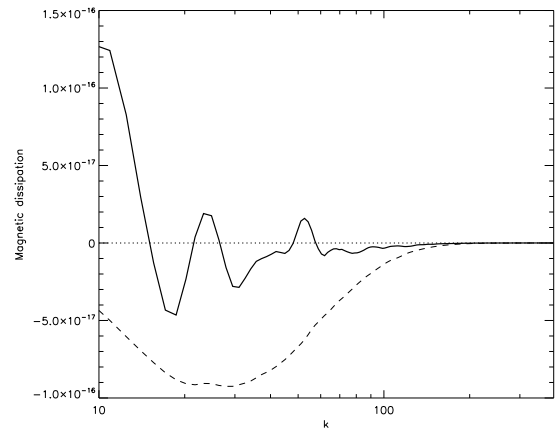}
\caption{The numerical dissipation rate
 is plotted vs $k$ using the solid line. 
This can be  compared with the 
  physical dissipation rate per unit volume
  indicated  using the dashed line. The latter has a
  larger amplitude than the former, indicating that the results
  are not strongly affected by numerical dissipation.}
\label{dissip_Re3125_Pr4}
\end{center}
\end{figure}

In paper I, we found that the dynamical properties of the turbulence
could be studied in Fourier space. The induction equation leads to a
balance between $5$ terms, describing
forcing by the mean and turbulent flow, transport to smaller scales,
compressibility and numerical dissipation. This gives rise to Eq.~(22)
of paper I. A similar balance involving $6$ terms can be obtained when
explicit resistivity is included. Using the same notation as in paper
I, we find
\begin{equation}
S + T_{bb} + T_{divv} + T_{bv} + D_{phys} + D_{num} = 0 \, .
\label{balance_eq}
\end{equation}
Here $D_{phys}$ describes physical dissipation per unit volume 
in ${\bf k}$ space and is defined by 
\begin{equation}
D_{phys}=\eta k^2 |\bb{B}(k)|^2 \, ,
\end{equation}
in which $k$ stands for the modulus of the wavenumber ${\bf k} \equiv
(k_x, k_y, k_z).$  As in paper I, 
$S$ describes how the background shear creates the
y component of the magnetic field, $T_{bb}$ is a term that accounts
for magnetic energy transfer toward smaller scales, $T_{divv}$ is due to
compressibility, $T_{bv}$ describes how magnetic field is created
due to field line stretching by the turbulent flow
and $D_{num}$ describes numerical dissipation.

The variation
of the first five terms of Eq.~(\ref{balance_eq}) with $k$ is plotted
in figure~\ref{transfer_tot_Re3125_Pr4} (to compute these curves, the
simulation data were averaged in time using $100$ snapshots spanning
$400$ orbits). $S$ is shown using the upper
solid line, the dashed line corresponds to $T_{bb}$ while $T_{divv}$,
$T_{bv}$ and $D_{phys}$ are respectively represented using the dotted
line, dotted--dashed line and lower solid line respectively.
As in paper I, the above quantities are Fourier transforms
evaluated with $k_y =0$ that are then averaged over
the  circle $k_x^2+k_z^2= {\rm constant.}$

 In agreement with the
results presented in paper I for simulations performed without
explicit dissipation, figure~\ref{transfer_tot_Re3125_Pr4} shows that
the only always  positive term is $S$, which is simply due to the fact that
toroidal magnetic energy is created by the mean background shear.
As in paper I, we next turn to the poloidal part of
Eq.~(\ref{balance_eq}). This is done by considering only the
components of the induction equation for the  poloidal
part $\bb{B}_p=(B_x,0,B_z)$ of the magnetic field $\bb{B}$. In that
case $S$ vanishes. The variation with $k$ of the four remaining terms
($T_{bb}$, $T_{bv}$, $T_{div}$ and $D_{phys}$) is shown on
figure~\ref{transfer_polo_Re3125_Pr4} using the same
conventions as in Fig.~\ref{transfer_tot_Re3125_Pr4}. Again, the
results we obtained are qualitatively similar
to the results of paper I. Poloidal magnetic energy is created on a
large range of scales by field line stretching, as indicated by the
fact that $T_{bv}$ is positive. The other terms are negative and
describe transport to smaller scales and dissipation. For the
simulation to be converged, physical dissipation need to be
larger than numerical dissipation. The interest of this Fourier
analysis is to provide a means to test this condition.
 Indeed, numerical
dissipation can be computed as minus the sum of the four terms plotted
on figure~\ref{transfer_polo_Re3125_Pr4}. Its variation with $k$ is
shown in figure~\ref{dissip_Re3125_Pr4} with the solid line
and compared to $D_{phys},$ represented using a dashed line. 
At small
scales ($k$ larger than $30$), numerical dissipation is clearly
smaller than physical dissipation. For the smallest $k$, large amplitude
fluctuations around zero are observed. They are due to poorer
statistics that prevent the procedure from converging everywhere.
Nevertheless, figure~\ref{dissip_Re3125_Pr4} provides confidence that
the dissipation is largely physical in model 128Re3125Pm4. This is
also consistent with
the results of paper I, which estimated a numerical magnetic Reynolds
number of the order of $30000$ at that resolution, significantly larger than
the value $Re_M=12500$ used in the present model (although we add a
note of caution that this estimated value depends on the flow
itself and could thus be slightly different here).

\subsection{Code comparison}
\label{code_compar_sec}

\begin{table*}[t]
\begin{center}\begin{tabular}{@{}cccccccc}\hline\hline
Code & Resolution & Box size & $\alpha_{Rey}$ & $\alpha_{Max}$ & $\alpha$ \\
\hline\hline
ZEUS & $(128,200,128)$ & $(H,\pi H,H)$ & $1.6 \times 10^{-3}$ & $7.4
\times 10^{-3}$ & $9.1 \times 10^{-3}$ \\
NIRVANA & $(128,200,128)$ & $(H,\pi H,H)$ & $1.7 \times 10^{-3}$ & $7.8
\times 10^{-3}$ & $9.5 \times 10^{-3}$ \\
SPECTRAL CODE & $(64,128,64)$ & $(H,\pi H,H)$ & $1.4 \times 10^{-3}$ & $9.4
\times 10^{-3}$ & $1.1 \times 10^{-2}$ \\
PENCIL CODE & $(128,200,128)$ & $(H,\pi H,H)$ & $1.4 \times 10^{-3}$ & $6.8
\times 10^{-3}$ & $8.2 \times 10^{-3}$ \\
\hline\hline
\end{tabular}
\caption{Details of the runs performed using different codes as a way
  of checking the results for model 128Re3125Pm4. The
  first three columns respectively describe the code used, the
  resolution $(N_x,N_y,N_z)$ and the box size $(L_x,L_y,L_z)$. The
  last three columns summarise the outcome  by giving the
  time averaged values of $\alpha_{Rey}$, $\alpha_{Max}$ and
  $\alpha$ (averaged between $t=40$ and the end of each
  simulation). Note that the first line simply recalls the results of
  model 128Re3125Pm4 which can also be found in
  Table~\ref{modes_properties_dissip}. All models have $Re=3125$ and
  $Pm=4$.}
\label{code_comparison}
\end{center}
\end{table*}

\begin{figure*}
\begin{center}
\includegraphics[scale=0.33]{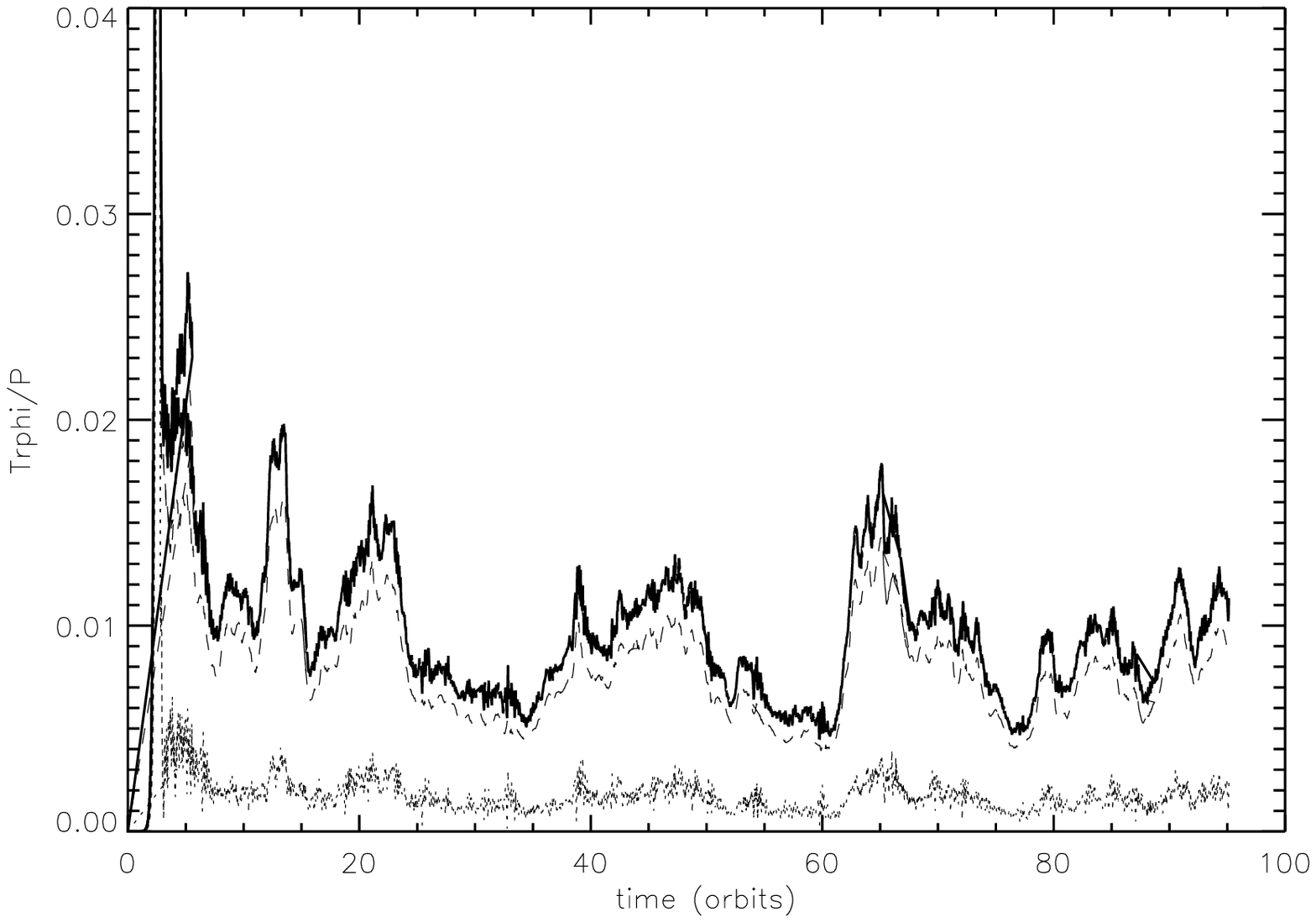}
\includegraphics[scale=0.33]{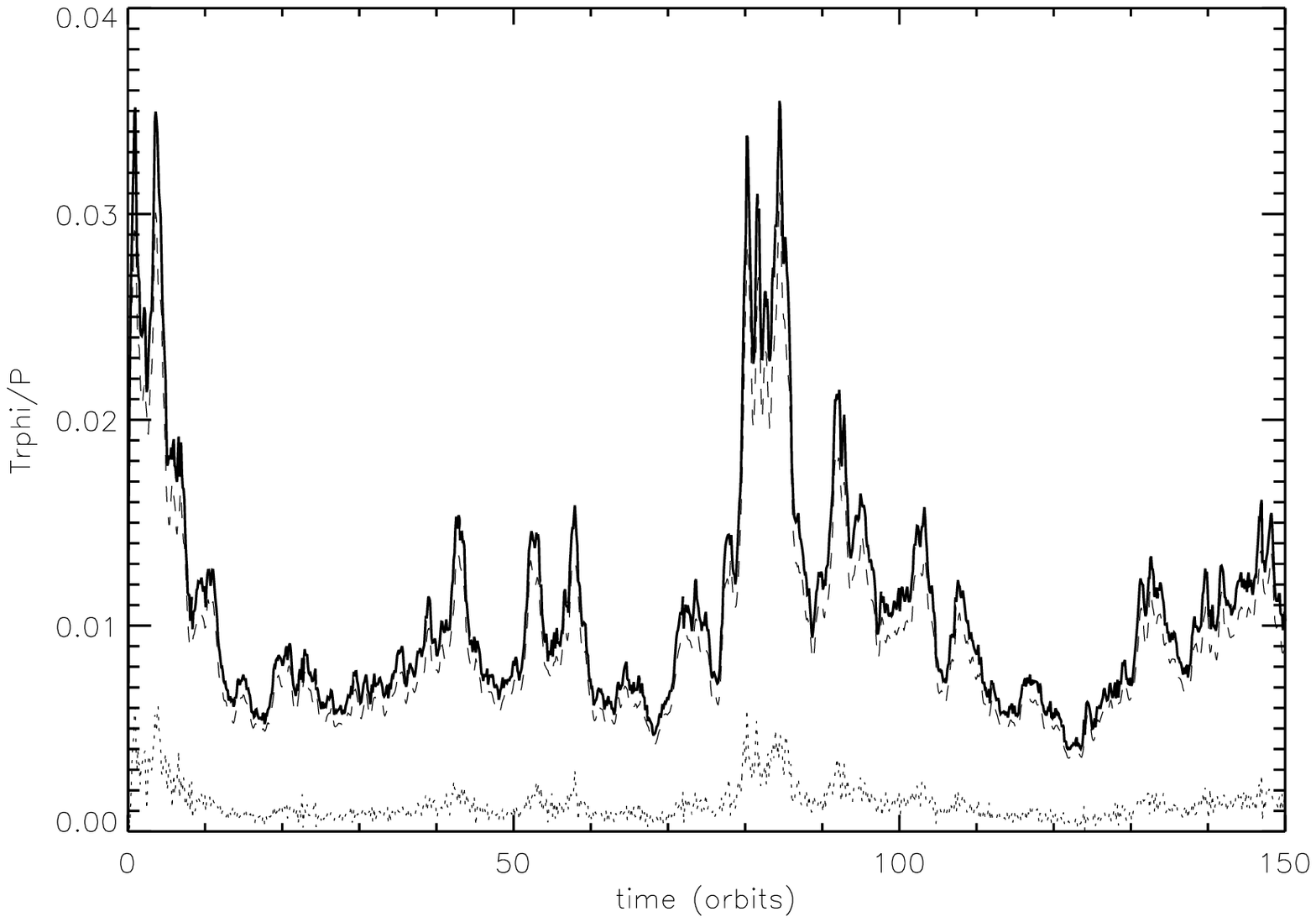}
\includegraphics[scale=0.33]{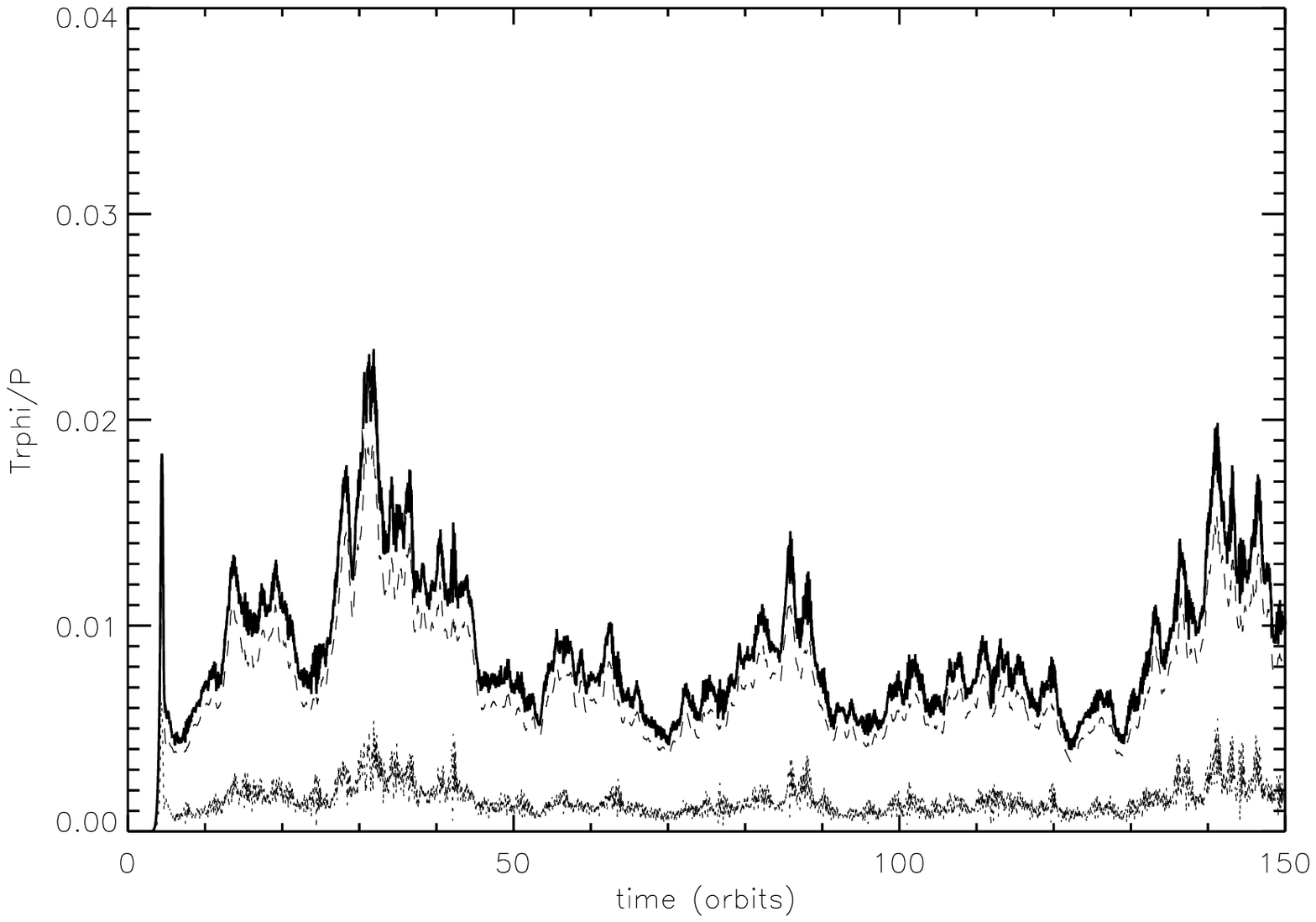}
\caption{Results obtained with NIRVANA ({\it left panel}), a spectral
  code ({\it middle panel}) and the PENCIL code ({\it right panel}) for
  the comparison case. All panels display the time history of
  $\alpha_{Rey}$, $\alpha_{Max}$ and $\alpha$ using the same conventions as
  figure~\ref{fiducial_transport}, with which they should be
  compared. Good agreement is found for model 128Re3125Pm4
  regardless of the numerical scheme used.}
\label{code_compar}
\end{center}
\end{figure*}

In order to gain further confidence that the results for model
128Re3125Pm4 are not strongly affected by numerical dissipation or
  numerical details such as, for example, the boundary conditions, we
reproduced that simulation using three other codes: NIRVANA
\citep{ziegler&yorke97}, a spectral code \citep{lesur&longaretti07}
and the PENCIL code \citep{brandenburg&dobler02}. All include the
same diffusion coefficients yielding $Re=3125$ and $Pm=4$.

NIRVANA is a finite difference code that uses the same algorithm as
ZEUS but was developed independently. It has been used 
frequently in the
past to study various problems involving MHD turbulence in the
shearing box \citep{pap&nelson03b,fromang&pap06}. The
  implementation of the shearing box boundary conditions is identical
  to that of ZEUS. The net magnetic fluxes introduced in the
  computational domain during the simulation are therefore of the same
  order as for ZEUS.

The Pencil Code uses sixth order central finite differences in space
and a third order Runge-Kutta solver for time integration.
In addition to the standard diffusion coefficients (resistivity and
viscosity), a sixth order hyper diffusion operator is used in the
continuity equation. Solenoidality is ensured by evolving the magnetic
vector potential. The boundary conditions are enforced directly on
  that potential using a six order polynomial
  interpolation. We monitored the accumulation of magnetic flux in the three
  directions and found effective $\beta$ of the order of $10^{19}$ in
  the radial direction and $10^{16}$ in the azimutal and vertical
  direction. The box size and resolution in this simulation
  are the same as in model 128Re3125Pm4 computed with ZEUS and the
  model was run for $150$ orbits.

The spectral code is based on a 3D Fourier expansion of the
resistive-MHD equations in the incompressible limit. It uses a
pseudo-spectral method to compute non linear interactions and is based
on the "3/2" rule to avoid aliasing. The flow is computed in the
sheared frame and a remap procedure is used, as described by
\citet{umurhan&regev04}. Since these routines are the main source of numerical
dissipation, the global energy budget is evaluated and we check that
viscosity and resistivity are responsible for more than 97\% of the
total dissipation \citep[see][for a description of
the energy budget control]{lesur&longaretti05}. Therefore, the
numerical dissipation is still
present, but is kept at a very low level compared to physical
dissipation. Finally, the boundary conditions are purely periodic
  in the sheared coordinates and the magnetic fluxes are
  conserved to round--off error. During the simulation, the maximum effective
  $\beta$ we measured are of the order of $10^{22}$. 
This code has been successfully used for pure HD
\citep{lesur&longaretti05} and MHD-MRI \citep{lesur&longaretti07}
problems. In this paper, it uses a box size $(L_x,L_y,L_z)=(H,\pi H,H)$,
a resolution  $(N_x,N_y,N_z)=(64,128,64)$ and was run for $150$
orbits. Note that we
  decreased the number of grid points in the radial direction by a
  factor of two for this model compared with the set up used by the
  other codes. This is because the spectral code,
  being equivalent spatially to a 64th--128th order finite difference code,
  can use a coarser resolution than
  the other numerical schemes and still resolve the same dissipation legnths.

The results we obtained using these three codes are summarised with
the help of figure~\ref{code_compar}, which shows the time history of
the various stresses as obtained with NIRVANA ({\it left panel}), the
spectral code ({\it middle panel}) and the PENCIL code ({\it right
  panel}). In plotting the different curves, we used the same
conventions as in figure~\ref{fiducial_transport}, with which the
results should be compared. In general, good agreement is found
between the four models. Time averaged values of the stresses,
measured between $t=40$ and the end of the simulation, are computed
using these models. As indicated in 
table~\ref{code_comparison}, one finds
 $\alpha=9.3 \times 10^{-3}$, $1.1 \times 10^{-2}$ and $8.2
\times 10^{-3}$, when respectively using NIRVANA, the spectral code
and the PENCIL code. Given the large diversity of the numerical methods
that are used in these codes and the different implementation of
  the boundary conditions they employ, this small scatter is an
indication that the effect of numerical issues is very small in model
128Re3125Pm4. 
In particular, the low level of numerical dissipation (of the order of
$3 \%$) obtained in the spectral code combined with the
suggestion of figure~\ref{dissip_Re3125_Pr4} that physical resistive
dissipation dominates over numerical resistive dissipation in ZEUS both 
provide evidence that physical dissipation dominates over numerical
dissipation in the simulations. At the very least, the good agreement
between the four codes suggest that any residual numerical dissipation is
not large enough to influence the transport properties in any of these
models.

\section{The parameter space}
\label{parameter_sec}

Having shown that the results of model 128Re3125Pm4, computed using
ZEUS, are not strongly affected by numerical dissipation, we now turn
into a more systematic exploration of the parameter space. 

In doing so, we adopt a typical resolution $(N_x,N_y,N_z)=(128,200,128)$ for
all cases in which $Re_M$ and $Re$ are smaller than $12500$. 
Given the above analysis, it is reasonable to assume that the viscous and
resistive lengths are sufficiently resolved in those models for the
transport properties to be accurately computed. 

In the following, we will also present one model
having $Re=12500$ and $Pm=2$ (and therefore $Re_M=25000$). Using the
same resolution in that case
would result in numerical dissipation being of the same order as
physical dissipation, since we demonstrated in paper I that the
numerical magnetic Reynolds number is around $30000$ when using $128$
cells per scale height. 
In addition, the resistive length scale will be reduced by about $40$\%
compared to model 128Re12500Pm1, for which $Re=12500$ and
$Pm=1$.  Indeed, as argued by \citet{schekochihinetal04} for the kinematic
regime, the visous dissipation length and the resistive dissipation
length are related through $l_{\nu} \sim Pm^{1/2} l_{\eta}$ (this
relation, not taking rotation into account, is not strictly speaking
applicable, but may be indicative).
For both of these reasons, we used the more computationally
  demanding resolution
$(N_x,N_y,N_z)=(256,400,256)$ in this case which should be enough to
ensure that numerical dissipation will not strongly affect the
results. Remember also that we showed in paper I that the numerical
magnetic Reynolds number is of the order of $10^5$ for such a
resolution, well above the magnetic Reynolds number of that model,
which also gives confidence that numerical dissipation
should not be dominant in this case. Reasoning along the same
  lines, we also used the same high resolution for model 256Re25000Pm1, for
which $Re=25000$ and $Pm=1$.

\begin{figure}
\begin{center}
\includegraphics[scale=0.5]{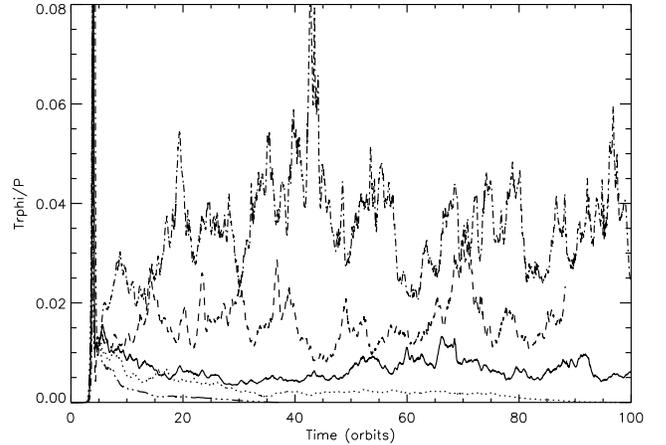}
\caption{Time history of $\alpha_{Max}$ in model 128Re800Pm16 ({\it
  dotted-dashed line}), 128Re1600Pm8 ({\it dashed line}), 128Re3125Pm4
  ({\it solid line}), 128Re6250Pm2 ({\it dotted line}) and 128Re12500Pm1
  ({\it dotted--dotted--dashed line}). In each cases, $Re_M=12500$,
  while $Pm$ is gradually decreased from $16$ for the top curve to $1$
  for the bottom curve. The results show an increase of activity when
  the Prandtl number increases. In addition, turbulent transport vanishes
  when $Pm \le 2$.}
\label{Pm_effect_Rem12500}
\end{center}
\end{figure}

\begin{figure}
\begin{center}
\includegraphics[scale=0.5]{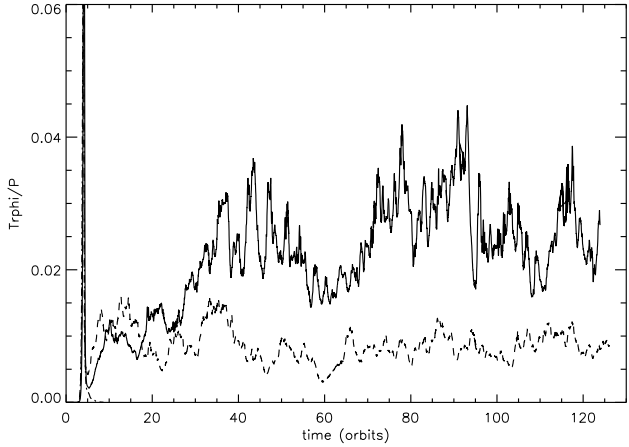}
\caption{Same as figure~\ref{Pm_effect_Rem12500}, but for models
  128Re800Pm8 ({\it solid line}), 128Re1600Pm4 ({\it dashed line}) and
  128Re3125Pm2 ({\it dotted-dashed line}), in which turbulence dies
  after about $5$ orbits. In each cases, $Re_M=6250$,
  while $Pm$ decreases from $8$ to $2$.}
\label{Pm_effect_Rem6250}
\end{center}
\end{figure}

\begin{figure}
\begin{center}
\includegraphics[scale=0.5]{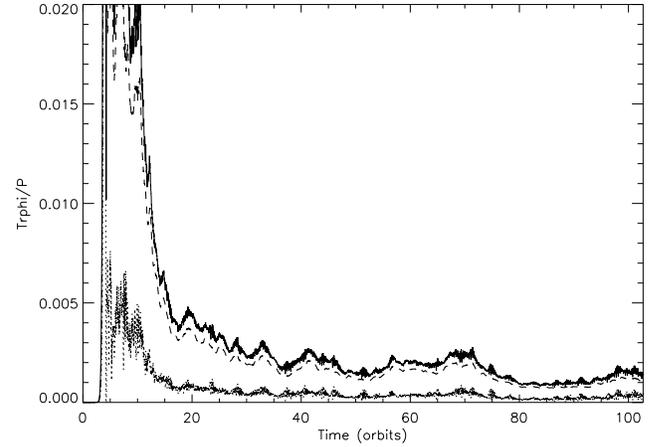}
\caption{Time history of $\alpha_{Re}$ ({\it dotted line}),
  $\alpha_{Max}$ ({\it dashed line}) and $\alpha$ ({\it solid line})
  for model 128Re12500Pm2. The rate of angular momentum transport,
  averaged from $t=40$ until the end of the simulation, gives
  $\alpha=1.6 \times 10^{-3}$.}
\label{hist_Re12500_Pm2}
\end{center}
\end{figure}

\begin{figure}
\begin{center}
\includegraphics[scale=0.5]{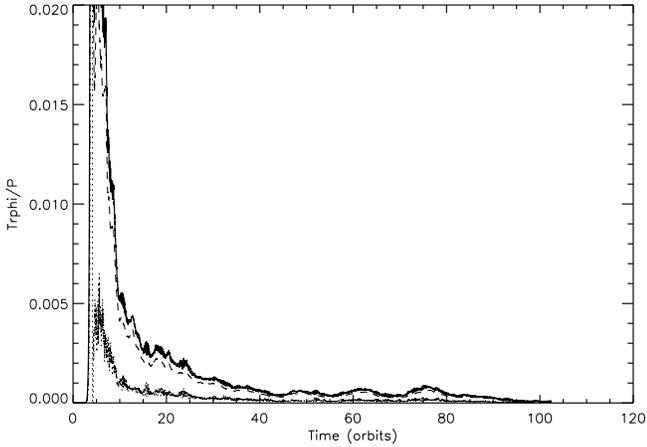}
\caption{Same as figure \ref{hist_Re12500_Pm2}, but
  for model 256Re25000Pm1. Turbulence decays after about
  100 orbits.}
\label{hist_Re2500_Pm1}
\end{center}
\end{figure}

\begin{figure*}
\begin{center}
\includegraphics[scale=0.7]{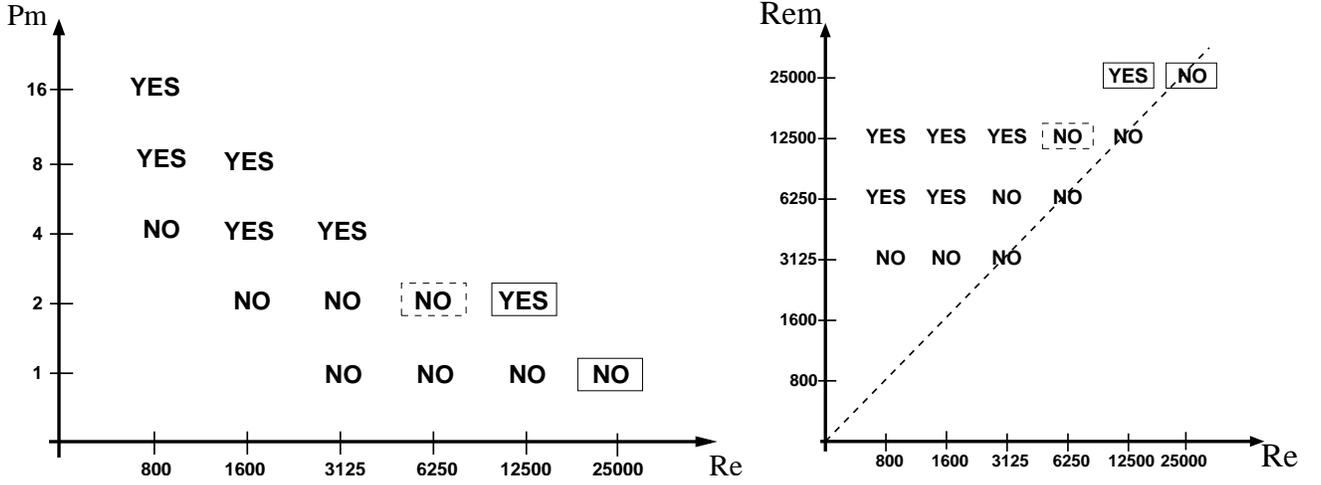}
\caption{Summary of the state (turbulent or not) of the flow in an
  $(Re,Pm)$ plane ({\it left panel}) and in an $(Re,Re_M)$ plane ({\it right
  panel}) for the models presented in this paper. In the later, the
  dashed line represents the $Pm=1$ case. On both panels, ``YES'' means
  that a non vanishing transport coefficient $\alpha$ was measure
  while ``NO'' means that MHD turbulence eventually decays:
  $\alpha=0$. All cases use a resolution
  $(N_x,N_y,N_z)=(128,200,128)$, except the models appearing in a
  solid squared box, for which the resolution was doubled. The model
  appearing in a dashed line squared box corresponds to the marginal
  model described in figure~\ref{Pm_effect_Rem12500}.}
\label{Re_Pm diagram}
\end{center}
\end{figure*}

The time averaged transport coefficients we measured in all the
simulations we performed are summarised in
Table~\ref{modes_properties_dissip} (for all models,
this average is done between $t=40$ and the end of the simulation). In
the present section, we now focus on a detailed examination of the
results. Figure~\ref{Pm_effect_Rem12500} shows the time history of
$\alpha_{Max}$ for all the simulations having $Re_M=12500$. They are
characterised by different values of the viscosity, in such a way that
$Pm=16$ ({\it
  dotted-dashed line}), $Pm=8$ ({\it dashed line}), $Pm=4$ ({\it solid
  line}), $Pm=2$ ({\it dotted line}) and $Pm=1$ ({\it
  dotted--dotted--dashed line}). It is obvious from these models that
angular momentum transport increases with the Prandtl
number, in agreement with the results of \citet{lesur&longaretti07}
obtained in the presence of a net vertical flux. In addition, MHD
turbulence is observed to die down in the last
two models. The critical Prandtl number below which turbulence is not
sustained is probably close to $Pm=2$. Indeed, model 128Re12500Pm2,
for which $Pm=2$, is seen to be marginal as it takes about $90$
orbits for the turbulence to decay. In the ``alive'' cases that display
turbulent activity, time averaged values of the total stress give
$\alpha=4.4 \times 10^{-2}$, $1.9 \times 10^{-2}$ and $9.1 \times
10^{-3}$, respectively when $Pm=16$, $8$ and $4$. For
fixed $Re_M$, this shows an almost linear scaling with viscosity. As
demonstrated with figure~\ref{Pm_effect_Rem6250}, the situation is
similar when using $Re_M=6250$. The three curves on this plot
correspond to $Pm=8$ ({\it solid line}), $Pm=4$ ({\it dashed line})
and $Pm=2$ ({\it dotted--dashed line}). Again, MHD turbulence disappear
when $Pm \le 2$ and increases with viscosity otherwise: $\alpha=9.7
\times 10^{-3}$ when $Pm=4$ and $\alpha=3.1 \times 10^{-2}$ when $Pm=8$.

Increasing $Re_M$ by a factor of two, we show in
figure~\ref{hist_Re12500_Pm2} the time history of
$\alpha_{Re}$ ({\it dotted line}), $\alpha_{Max}$ ({\it dashed line})
and $\alpha$ ({\it solid line}) for model 256Re12500Pm2, in which
$Re=12500$ and $Pm=2$. As shown in
Table~\ref{modes_properties_dissip}, MHD turbulence is sustained in
this case. Averaging $\alpha$ between $t=40$ and the end of the
simulation, we obtained $\alpha=1.6 \times 10^{-3}$. This is important
as MHD turbulence decays for all the other models we performed that
have $Pm=2$ and lower Reynolds numbers. Therefore, the results of
model 256Re12500Pm2 demonstrate that it is possible to obtain
sustained angular momentum at fairly low Prandtl number provided the
Reynolds number is large enough. However, model 256Re25000Pm1, in
  which $Re=25000$ and $Pm=1$, fails to show sustained turbulence for
  long times, as demonstrated in figure \ref{hist_Re2500_Pm1} in which
  the time history of the transport quantities is plotted. Although it
  takes about $100$ orbits for turbulence to decay, $\alpha$
  eventually decreases down to zero at the end of the
  simulation. Given how marginal this model seems to be, it seems
  plausible that a further increase of the Reynolds number will
  eventually lead to nonzero transport at $Pm=1$. The large resolution
needed, however, precludes such a simulation being run at the present
time.

The overall results of our simulations are summarised in
figure~\ref{Re_Pm diagram} which gives the state of the flow for each
models in an $(Re,Pm)$ plane ({\it left panel}) and in an $(Re,Re_M)$
plane ({\it right panel}). The flags ``YES'' means the disk is
turbulent, ``NO'' that turbulence was found to decay. The two cases
appearing in a solid squared
box have $Re_M=25000$ and use a resolution
$(N_x,N_y,N_z)=(256,400,256)$. The model appearing in a dashed squared 
box is the marginal case having $Re_M=12500$ and $Pm=2$ (see
figure~\ref{Pm_effect_Rem12500}). In general, this plot shows that MHD
turbulence is easier to obtain at large Prandtl number. When the disk is
turbulent, the results also suggest that angular momentum transport
increases with
$Pm$. None of the models having $Pm=1$ show any sign of activity,
although turbulence takes longer and longer to die as the Reynolds
number is increased. Taken together, these results demonstrate that
for any $Re$, there exists a critical Prandtl number $Pm_c$ below which
turbulence decays, while it is sustained when $Pm>Pm_c$. The results
presented in this paper suggest that $Pm_c$ decreases with
$Re$. However, there are not enough data
obtained from the current simulations to conclude that any asymptotic limit has
been reached or even to conclude to the existence of such a limit. It
is therefore not possible to extrapolate the behaviour of $Pm_c$ at
large $Re$.

\section{Discussion and conclusion}
\label{conclusion}

\subsection{Comparison with the results of paper I}

\begin{figure}
\begin{center}
\includegraphics[scale=0.45]{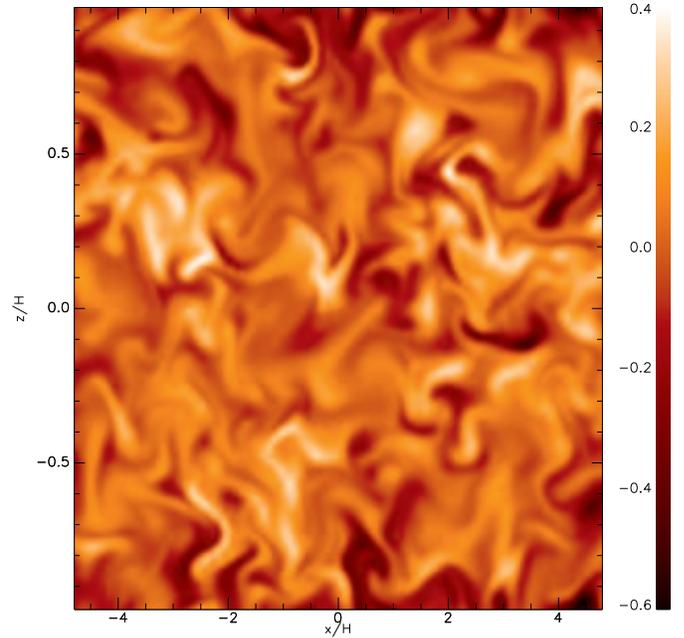}
\caption{Snapshots of $B_y$ in the $(x,z)$ plane at time $t=66$ in
  model 256Re12500Pm2. The structure of the flow and the typical
  length scale of the fluctuations are similar to that obtained in
  model STD128 in paper I (see the middle panel of figure 4 in paper I
  with which the present figure should be compared).}
\label{snap_256Re12500Pm2}
\end{center}
\end{figure}

In this paper, we studied the effect of finite dissipation
coefficients on the saturated level of MRI--driven MHD turbulence. One
of the important aspect of our results is that MHD turbulence can only
be sustained if the magnetic Prandtl number $Pm$ is larger than some
critical value $Pm_c$. For the limited range of Reynolds numbers that could be
probed given current day limitations in computing time, we found that
$Pm_c>1$. We want to stress here that these results are consistent
with the results we obtained in paper I. Indeed, we argued in section
5.1 of that paper that it is possible to derive a
``numerical''  magnetic Prandtl number when using ZEUS without
explicit diffusion coefficients and estimated its value, although very
uncertain, to be of order unity and probably somewhat larger than
one. The fact that MHD turbulence was found to be sustained in
these simulations is therefore consistent with the results of the
present paper, for which all of our ``alive'' case have $Pm$ larger
than unity. If ZEUS had had a numerical magnetic Prandtl number
smaller than
unity, we would predict from the present result that MHD turbulence
would decay when performing such simulations without explicit
dissipation coefficients. 

It is in fact possible to push the comparison between the results
of paper I and paper II a bit further. Indeed, we note that model
STD128 (presented in paper I) and model 256Re12500Pm2 (presented in
this paper) have similar time averaged value of $\alpha$: $2.2
\times 10^{-3}$ for the former and $1.6 \times 10^{-3}$ for the
later. This similar result is due to the fact that both model lie in
the same region of the $(Re,Pm)$ plane: for model 256Re12500Pm2,
$Re_M=25000$ and $Pm=2$. For model STD128, we estimated in paper I that
$Re_M$ is of the order of $30000$ while $Pm$ is of the order or
slightly larger
than one. This is why the results are similar. This is further illustrated
on figure~\ref{snap_256Re12500Pm2} which shows a snapshot of $B_y$ at
time $t=66$ in model 256Re12500Pm2. Note how similar it is to the
middle panel of figure~4 in paper I, which shows the same variable for
model STD128. Measuring the typical length scale $L_y(B_y)$ of the magnetic
structures in model 256Re12500Pm2 using Eq.~(9) of paper I, we found
a time averaged value $L_y(B_y)=0.045$, very close to the value $0.04$
we obtained
for model STD128. 

It is also possible to compare the results of model
STD64 of paper I to the results of the present paper. We recall here
that we
found the rate of angular momentum transfer in this model to be such
that $\alpha \sim 0.004$ when time averaged over the simulation. For
model STD64, we estimated in paper I that $Re_M \sim 10^4$ and a similar
value for the magnetic Prandtl number as for model STD128. This would
correspond to Reynolds number somewhat smaller than $10000$. In the
present paper, we found that $\alpha \sim 0.01$ for model
128Re3125Pm4, for which $Re=3125$ and $Pm=4$, while model
128Re3125Pm2, having $Re=3125$ and $Pm=2$ was shown to decay. It is
therefore tempting to identify model STD64 with a
model that would be intermediate between the last two cases. Using the
PENCIL code, we ran such a model, having $Re=3125$ and $Pm=3$, and
found $\alpha=0.007$ which is close to the result of model STD64. 

We want to stress, however, that it would be dangerous to push such
comparisons further than that. Indeed, we demonstrated in
paper I that numerical dissipation generally departs from a pure
Laplacian dissipation in ZEUS. Moreover, we stressed in paper I that
an accurate estimate for the magnetic Prandtl number is difficult to
obtain for a given simulation, as it depends on the nature of the flow
itself. A one to one comparison between the results of paper I and
paper II is therefore difficult to carry and may not bear much significance.

\subsection{Small scales}

The results of this paper together with paper I indicate the
importance of flow phenomena occurring at the smallest scales
available in a simulation, at least at currently feasible
resolutions. In fact the importance of small scales determined by the
transport coefficients is not unexpected when one considers previous
work on the maintenance of a kinematic magnetic dynamo.

Although a kinematic dynamo considers only the induction equation with
an imposed velocity field, some issues arising in that case may be
relevant, especially if one wishes to consider the likely behaviour of
turbulence driven by the MRI when the transport coefficients are
reduced to very small values.

If a dynamo is to be maintained in a domain such as a shearing box
with no net flux, one would expect that the magnitude of  a magnetic
field could be amplified from a small value through the action of some
realised velocity field. Furthermore if such an amplification occurs
within a specified time scale and for arbitrarily small resistivity,
it would be classified as a fast dynamo.  In the special case when the
imposed velocity field is stationary \citet{MoffProc85} have shown
that the field produced by such a dynamo must have a small spatial
scale determined by the resistivity. A well known example of this type
is generated by the so called  'ABC' flow \citep[see][and
  references therein]{teyssieretal06}. This example also shows that certain
quantities such has the growth rate of the dynamo do not have a simple
dependence on magnetic Reynolds number when that is relatively small
and thus caution should be exercised in making any simple
extrapolation.

 Although the case of a steady state velocity field  is rather
special, the result can be very easily  seen to  hold more generally
for the case when the    magnetic field is presumed to have net
helicity. The magnetic helicity ${\cal H},$ which may be regarded as a
topological quantity measuring the degree of entanglement or
knottedness of the field,  is defined by
 \begin{equation}
{\cal H}= \int_V \bb{A} \cdot \bb{B} dV,
\end{equation}
where ${\bf A}$ is the vector potential and the integral is over the
whole box.  In our case the boundary conditions ensure ${\cal H}$ is
gauge invariant.  In ideal MHD ${\cal H}$ is conserved and so cannot
grow by dynamo  action. When the resistivity is non zero, the
induction equation implies that \citep{MoffProc85}
 \begin{equation}
{d {\cal H}\over dt} = -2\int_V \eta (\nabla\times {\bf
A})\cdot(\nabla\times {\bf B})dV.\label{helicity}
\end{equation}
Using the Schwartz inequality, this implies that

 \begin{equation}
\left |{1\over {\cal H}}{d {\cal H}\over dt}\right | \le  2\eta_{max}
|{\cal K}|l^{-2},
\end{equation}
where $\eta_{max}$ is the maximum value of $\eta,$ ${\cal K}$ is the
inverse of the relative helicity given by
 \begin{equation}
{\cal K} = \sqrt{\int_V {\bf A}^2dV\int_V {\bf B}^2dV}/{\cal H}
\end{equation}
and the length scale $l$ is defined through
 \begin{equation}
l^{-4} = {\int_V |\nabla\times {\bf A}|^2dV \int_V |\nabla\times {\bf
B}|^2dV\over  \int_V {\bf A}^2dV\int_V {\bf B}^2dV}.
\end{equation}
From equation (\ref{helicity}) one can argue that for any field with
non zero relative helicity to grow in a finite time, the scale length
$l$ must decrease to arbitrarily small scales as the transport
coefficients are decreased to small values.

Although the above problems are not directly dealing with the MRI, and
the discussion is  by no means comprehensive, it is suggestive in
indicating that significant flow structures  plausibly  occur at small
scales as the transport coefficients are decreased and also that some
features may not vary monotonically   as a function of Reynolds number.

 Another important feature apparent in the results presented here is the
 increase of turbulent activity with Prandtl number at fixed Reynolds
 number.  This  is difficult to quantify but may be connected to the
 increasing  inhibition of reconnection by moving field lines together
 when the viscosity increases.

In this context   it is important to emphasize that the simulations discussed here were carried out
using an isothermal equation of state. However, there can be situations where
a thermal diffusivity needs to be considered in addition to viscosity and resistivity.
The isothermal approximation should be reasonable as long as any length scale
introduce by heat diffusion is significantly longer than
the resistive or viscous scales. When this length scale becomes comparable
to the others we would expect the velocity and pressure fields to be affected on this scale.
This too may affect reconnection rates, the operation of a small scale dynamo
and a consequent inverse cascade. 
But the necessity of small scales as argued above depends on
a balance determined by the induction equation and is not affected.

\subsection{Towards an asymptotic regime for the zero net flux shearing box}

\begin{figure}
\begin{center}
\includegraphics[scale=0.6]{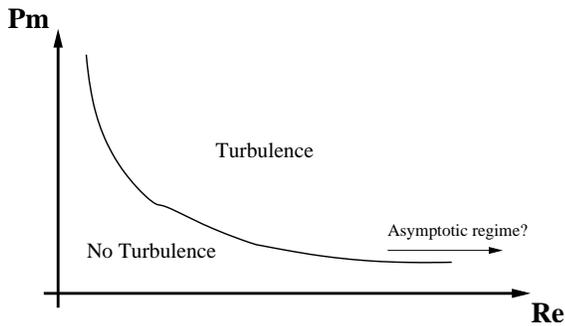}
\caption{Cartoon showing the variation of the critical Prandtl number
  $Pm_c$, above which MHD turbulence is sustained as a function of the
  Reynolds number. As simulations have not attained an asymptotic
  regime,  qualitative changes of behaviour cannot be ruled out if
  $Re$  were to be increased further.}
\label{critical_pm}
\end{center}
\end{figure}

As we mentioned above, we have  demonstrated in the present paper that
for each value of the Reynolds number, there exists a critical Prandtl
number $Pm_c$ below which turbulence decays. The variation of $Pm_c$
with $Re$ is sketched on figure~\ref{critical_pm}. Given this plot,
one may want to extrapolate the behaviour of $Pm_c$ at large $Re$. We
want to stress that such an extrapolation cannot be made based on  our
results: we did not  reach  an asymptotic regime within the parameter
regime we were able to study.  This will require further simulations,
performed at larger Reynolds number. Such calculations will have to be
performed at resolutions $(512,800,512)$ and  higher and will be
extremely demanding computationally. At the present time, any of the
following behaviour should be considered possible: $Pm_c$ could
monotonically decrease to zero, tend to a finite value or reach a
minimum   increasing again  for larger $Re$.

However, it is interesting to note in this context
 that recent work on fluctuating dynamos (driven by external forcing)
indicate that they can be maintained at small $Pm$ but 
increasing values of $Re$ are required as this decreases
\citep{boldyrev&cattaneo04,Ponty05,Schekochihin07} exacly as one would
 extrapolate
using Figure \ref{critical_pm}. We also point out that
these results, as do those based on kinematic dynamos,
 indicate the importance of dynamo generated fields  on  small
scales and also that  these may in turn act as a source for field
on larger scales. Clearly simulations at  much higher resolution  
are required to investigate the existence of such a regime in our case.

Another question that cannot be answered  yet is whether $\alpha$
attains a finite limit at large Reynolds number for a given Prandtl
number. The answer to this question would also require  the resolution
to be increased in order to extend the   domain that can be probed in
the $(Re,Pm)$ plane.  Both issues are astrophysically important and
need to be addressed in the future as soon as the computing resources
become more readily accessible to the community.

\subsection{Future work}

It is also important to stress that the work in this paper applies to
an extremely simple set up of an unstratified shearing box with zero net
flux.  The
reason for focusing on this case was that it has been considered that
it could offer the possibility of a dynamo in the local limit, this
being  independent of exterior boundary conditions and imposed
magnetic fields.

The work here indicates  the possibility of the importance of small
 scales and only modest angular momentum transport in this limit.  It
 therefore also emphasises the need for studies of more general
 configurations which take account of  stratification,  global
 boundary conditions and imposed fields.  There is no reason to
 suppose that consideration of transport coefficients and  small scale
 phenomena are not important in these cases also so this should be an
 important area for future investigations.

\section*{ACKNOWLEDGMENTS}

We thank Gordon Ogilvie, Fran{\c c}ois Rincon and Alex Schekochihin for
useful discussions. The simulations presented in this paper were
performed using the Cambridge High Performance Computer Cluster
Darwin, the UK Astrophysical Fluids Facility (UKAFF), the Institut du
D\'eveloppement et des Resources en Informatique Scientifique (IDRIS),
the Grenoble observatory computing center (SCCI) and the Danish Center
for Scientific Computing (DCSC). We thank the referee, Jim Stone, for
helpful suggestions that significantly improved the paper.

\bibliographystyle{aa}
\bibliography{7943}

\newcommand{\noopsort}[1]{}
\begin{thebibliography}{22}
\expandafter\ifx\csname natexlab\endcsname\relax\def\natexlab#1{#1}\fi

\bibitem[{Balbus \& Hawley(1998)}]{balbus&hawley98}
Balbus, S. \& Hawley, J. 1998, Rev.Mod.Phys., 70, 1

\bibitem[{{Boldyrev} \& {Cattaneo}(2004)}]{boldyrev&cattaneo04}
{Boldyrev}, S. \& {Cattaneo}, F. 2004, Physical Review Letters, 92, 144501

\bibitem[{{Brandenburg} \& {Dobler}(2002)}]{brandenburg&dobler02}
{Brandenburg}, A. \& {Dobler}, W. 2002, Computer Physics Communications, 147,
  471

\bibitem[{{Fleming} {et~al.}(2000){Fleming}, {Stone}, \&
  {Hawley}}]{flemingetal00}
{Fleming}, T.~P., {Stone}, J.~M., \& {Hawley}, J.~F. 2000, ApJ, 530, 464

\bibitem[{{Fromang} \& {Papaloizou}(2006)}]{fromang&pap06}
{Fromang}, S. \& {Papaloizou}, J. 2006, A\&A, 452, 751

\bibitem[{{Fromang} \& {Papaloizou}(2007)}]{Fropap07}
{Fromang}, S. \& {Papaloizou}, J.~C.~B. 2007, A\&A, submitted

\bibitem[{{Goldreich} \& {Lynden-Bell}(1965)}]{goldreich&lyndenbell65}
{Goldreich}, P. \& {Lynden-Bell}, D. 1965, MNRAS, 130, 125

\bibitem[{Hawley \& Stone(1995)}]{hawley&stone95}
Hawley, J. \& Stone, J. 1995, Comput. Phys. Commun., 89, 127

\bibitem[{{Hawley} {et~al.}(1995){Hawley}, {Gammie}, \&
  {Balbus}}]{hawleyetal95}
{Hawley}, J.~F., {Gammie}, C.~F., \& {Balbus}, S.~A. 1995, ApJ, 440, 742

\bibitem[{{Hawley} {et~al.}(1996){Hawley}, {Gammie}, \&
  {Balbus}}]{hawleyetal96}
{Hawley}, J.~F., {Gammie}, C.~F., \& {Balbus}, S.~A. 1996, ApJ, 464, 690

\bibitem[{{Iskakov} {et~al.}(2007){Iskakov}, {Schekochihin}, {Cowley},
  {McWilliams}, \& {Proctor}}]{Schekochihin07}
{Iskakov}, A., {Schekochihin}, A., {Cowley}, S., {McWilliams}, J.~C., \&
  {Proctor}, M. R.~E. 2007, Physical Review Letters, 98, 208501

\bibitem[{{Landau} \& {Lifshitz}(1959)}]{landau&lifchitz59}
{Landau}, L.~D. \& {Lifshitz}, E.~M. 1959, {Fluid mechanics} (Course of
  theoretical physics, Oxford: Pergamon Press, 1959)

\bibitem[{{Lesur} \& {Longaretti}(2005)}]{lesur&longaretti05}
{Lesur}, G. \& {Longaretti}, P.-Y. 2005, \aap, 444, 25

\bibitem[{{Lesur} \& {Longaretti}(2007)}]{lesur&longaretti07}
{Lesur}, G. \& {Longaretti}, P.-Y. 2007, \mnras, 378, 1471

\bibitem[{{Moffatt} \& {Proctor}(1985)}]{MoffProc85}
{Moffatt}, H.~K. \& {Proctor}, M.~R.~E. 1985, JFM, 154, 493

\bibitem[{{Papaloizou} {et~al.}(2004){Papaloizou}, {Nelson}, \&
  {Snellgrove}}]{pap&nelson03b}
{Papaloizou}, J.~C.~B., {Nelson}, R.~P., \& {Snellgrove}, M.~D. 2004, MNRAS,
  350, 829

\bibitem[{{Ponty} {et~al.}(2005){Ponty}, {Mininni}, {Montgomery}, {Pinton}, \&
  Pouquet}]{Ponty05}
{Ponty}, Y., {Mininni}, P., {Montgomery}, D., {Pinton}, J.-F.~Politano, H., \&
  Pouquet, A. 2005, Physical Review Letters, 94, 164502

\bibitem[{{Sano} {et~al.}(2004){Sano}, {Inutsuka}, {Turner}, \&
  {Stone}}]{sanoetal04}
{Sano}, T., {Inutsuka}, S., {Turner}, N.~J., \& {Stone}, J.~M. 2004, ApJ, 605,
  321

\bibitem[{{Schekochihin} {et~al.}(2004){Schekochihin}, {Cowley}, {Taylor},
  {Maron}, \& {McWilliams}}]{schekochihinetal04}
{Schekochihin}, A.~A., {Cowley}, S.~C., {Taylor}, S.~F., {Maron}, J.~L., \&
  {McWilliams}, J.~C. 2004, \apj, 612, 276

\bibitem[{{Teyssier} {et~al.}(2006){Teyssier}, {Fromang}, \&
  {Dormy}}]{teyssieretal06}
{Teyssier}, R., {Fromang}, S., \& {Dormy}, E. 2006, Journal of Computational
  Physics, 218, 44

\bibitem[{{Umurhan} \& {Regev}(2004)}]{umurhan&regev04}
{Umurhan}, O.~M. \& {Regev}, O. 2004, \aap, 427, 855

\bibitem[{{Ziegler} \& {Yorke}(1997)}]{ziegler&yorke97}
{Ziegler}, U. \& {Yorke}, H.~W. 1997, Computer Physics Communications, 101, 54

\end{thebibliography}

\end{document}